\documentclass[aps,preprint,superscriptaddress,floatfix]{revtex4-1}
\usepackage[pdftex]{graphicx}% Include figure files
\usepackage{dcolumn}% Align table columns on decimal point
\usepackage{bm}% bold math
\usepackage{amssymb}
\usepackage[utf8]{inputenc}
\usepackage{color} % add this to use soul
\usepackage{soul}  % add this to use \hl - highlighting

\begin{document}

\title{Phase diagram of diblock copolymer melt in dimension d=5}

\author{M. Dziecielski}
\author{K. Lewandowski}
 \author{M. Banaszak}
\email[]{mbanasz@amu.edu.pl   Corresponding Author}

\affiliation{ Faculty of Physics,
A. Mickiewicz University \\
ul. Umultowska 85,
61-614 Poznan,
Poland}

\date{\today}

\begin{abstract}
Using self-consistent field theory (SCFT) in spherical unit cells of various dimensionalities, $D$,   a  phase diagram of a diblock, $A$-b-$B$, is calculated  in 5 dimensional space, $d$=5.
This is an extension of a previuos work for $d$=4.
The phase diagram is parameterized by the chain composition, $f$, and  incompatibility between $A$ and $B$,  quantified by the product $\chi N $.
We predict 5 stable nanophases:
layers, cylinders, 3$D$ spherical cells,  4$D$  spherical cells, and 5$D$ spherical cells.   In the strong segregation limit, that is for large $\chi N$, the order-order
transition compositions are determined by the strong segregation theory (SST) in its simplest form.
While the predictions of the SST theory  are close   to the
corresponding SCFT extrapolations for $d=4$, the extrapolations for
$d=5$ significantly differ from them. 
We find that 
 the $S_5$ nanophase is stable in a narrow strip between
ordered $S_4$ nanophase and the disordered phase.
The calculated order-disorder transition lines depend weakly on $d$, as expected.

%Unit Cell Approximation`

\end{abstract}

\pacs{}
%\keywords{}

\maketitle

\section{Introduction}

Diblock copolymer  (DBC), $A$-b-$B$,  melts consist of 2 types of segments, $A$ and $B$,  arranged in 2 
corresponding blocks.  Those melts can  self-assemble in $3d$ into various spatially-ordered nanophases,
such as layers, ($L$), hexagonally packed cylinders, ($C$), gyroid nanostructures, ($G$), with the $Ia\overline{3}d$ symmetry, and cubically packed (either body-centered or closely packed)  spherical cells ($S$),
depending on the chain composition, $f$ ($f$ is the fraction of $A$-segments; $1-f$ is the fraction of $B$-segments), degree of polymerization (number of segments), $N$, and the temperature-related $\chi$ parameter \cite{hamley2004, fredrickson2006b}. 
Recently, an additional  $O^{70}$-phase has been reported \cite{bailey02, takenaka07}, but it is stable in a very small region of the phase diagram.
Those nanophases can be transformed into a disordered phase, for example, upon heating. It is of great interest to determine a phase diagram of such melts exhibiting  order-disorder transition (ODT) lines, also referred to  as binodals of microphase separation transition (MST), and order-order transition (OOT) lines. 
This task has been largely achieved for 3-dimensional (bulk) diblock melts by accumulating results from numerous
experimental and theoretical studies \cite{lei80:9, ban92, vav92,  mat94, whit96, mat2001j, lennon2008, taniguchi09}, also for 2$d$ diblock copolymer melts \cite{hamley2009}.

The L, C, and S nanophases are known as classical, whereas  G and $O^{70}$  nanonophases are
referred to as non-classical, or sometimes complex.
The Wigner-Seitz cell of a classical phase can be approximated by $D$-dimensional sphere, $S_D$, both in 
the real $\vec{r}$-space and the reciprocal $\vec{k}$-space. Within this approximation, known as 
Unit Cell Approximation (UCA), the L, C, S nanophases correspond to $S_1$, $S_2$, and $S_3$, respectively, 
and the spacial distribution of chain segments can be mapped with a single radial variable, $r$, as shown in Table \ref{tab_tab4}. 
The classical phases can be easily generalized to higher dimensions, in particular 
for $d=5$ we can have 5 nanophases $S_D$, with dimensionality, $D$, ranging from 1 to 5.

It is interesting that a  mean-field (MF) theory applied to copolymer melts \cite{mat94, matbook, coch06}, known as the Self-Consistent Field Theory (SCFT),
 is  successful in
predicting diblock phase diagrams resembling the experimental ones, as shown, for example, in ref \cite{kha95:5}.
 The SCFT approach is
based on the assumption that coarse-grained polymer chains in dense melts  are Gaussian  (Flory's theorem \cite{deg79}), and  on the MF approximation which selects the dominant contribution in the appropriate partition function, thus neglecting fluctuations.

Because, in the MF theories, it is sufficient to know the
composition, $f$, and the product $\chi N$ in order to foresee the
nanophase \cite{lei80:9, mat02, coch06}, 
the diblock phase diagram can be mapped in ($f$, $\chi N$)-plane.
The MF theories exist in many variations, both in real space ($ \vec{r}$-space)\cite{ban92, vav92, whit96}  and Fourier space ($\vec{k}$-space) \cite{mat94, fredrickson2006b}.

\begin{table}\label{tab_tab4}
\caption{Unit cell equations of $D$-dimensionality; equations
are supplemented with the unconstrained variables for corresponding $d$'s (2, 3, 4 and 5); * indicates
the absence of unconstrained variables; imp  indicates that the nanophase for this $d$ is impossible}
\centering
% use packages: array
\begin{tabular}{|c|c|c|c|c|c|c|c|} \hline 
 $D$ & nanophase & cell equation & d=2 & d=3 & d=4 & d=5  &radial coordinate \\  \hline 
1 & L ($S_1$) & $x^2 <R^2$ & $y$ & $y$, $z$ &$y$, $z$, $t$& $y$, $z$, $t$, $v$ & $r=|x|$ \\  \hline
2 & C ($S_2$) & $x^2 + y^2 < R^2$ & * & $z$  & $z$, $t$  & $z$, $t$, $v$ & $r=\sqrt{x^2+y^2}$ \\  \hline
3 & $S_3$& $x^2+y^2+z^2<R^2$ & imp  & * & $t$ & $t$, $v$ &$ r= \sqrt{x^2 +y^2 +z^2}$  \\ \hline 
4 & $S_4$& $x^2+y^2+z^2+t^2<R^2$ &imp & imp  & * &  $v$  & $r= \sqrt{x^2 +y^2 +z^2 +t^2}$  \\ \hline 
5 & $S_5$& $x^2+y^2+z^2+t^2+v^2<R^2$ &imp & imp  & imp & *  & $r= \sqrt{x^2 +y^2 +z^2 +t^2+v^2}$  \\ \hline 
\end{tabular}
\end{table}

In addition, we intend to compare the phase boundaries calculated by the SCFT (and extrapolated to the strong segregation limit) with 
the strong segregation theory (SST)  for diblock melts, developed by Semenov \cite{sem85},
in which 
the free energy of the nanophase has three contributions,
the interfacial tension  and the stretching (of entropic origin) energies of the A and B blocks.
These energies can  be approximated
by simple expressions, allowing the calculation of the OOT compositions in the SST.
The main goal of this paper is to construct a phase diagram of a copolymer melt  for $d=5$, applying 
the SCFT  method with the UCA in $r$-space,  as presented in \cite{ban92, vav92, mat94}.
Specifically, we intend  to determine the area in  ($f$, $\chi N$)-space, in  which the $S_5$ phase is stable, 
by varying both the radius, $R$,  of the unit cell and the dimensionality, $D$.

In previous work\cite{previous} we calculated the phase diagram of the diblock
copolymer melt for $d=4$, and  managed  to aswer the following questions:
\begin{enumerate}
 \item is the $S_4$ nanophase  stable?
\item what is the sequence of nanophases, upon changing $f$?
\item are the binodals (ODT lines)  shifted as we vary $d$ from 1 to 4?
\item what are the strong segregation limits of the OOT lines for $d=4$? 
\end{enumerate}
The answers were as follows:
\begin{enumerate}
 \item  the nanophase $S_4$ is stable within a relatively narrow strip between  the $S_3$ nanophase and the disordered phase,
\item  the sequence of nanophases appropriate for the UCA in 3$d$ is preserved, starting from $f=1/2$,  $L$, $C$, $S_3$, and 
there is an additional
$S_4$ nanophase in 4$d$,
\item  the ODT binodals depend weakly on $d$, and they are shifted as  $d$ is varied,
\item the  SST compositions, $f_{L/C}$,  $f_{C/S_3}$, and $f_{S_3/S_4}$ 
are close   to the
corresponding extrapolations from the self-consistent field theory.
\end{enumerate}

In this work, similarly as in ref \onlinecite{previous}, the following questions are  posed:
\begin{enumerate}
 \item is the $S_5$ nanophase  stable?
\item what is the position of the $S_5$ nanophase in sequence of phases, upon changing $f$?
\item are the binodals (ODT lines)  shifted as we vary $d$ from 4 to 5?
\item what are the strong segregation limits of the OOT lines for $d$=5? 
\end{enumerate}

\newpage
\section{Method}
The incompressible copolymer melt is modeled as a collection of $n$ diblock  chains confined in volume $V$. Each chain, labeled $\alpha = 1, 2, \dots , n$, can take any
Gaussian configuration (in accordance with the Flory's Theorem \cite{deg79}) parameterized from $s$=0 to $s$=$f$ for $A$-segments, and from $s$=$f$ to $s$=1 for $B$-segments. Up to a multiplicative constant, the partition 
function for a \textit{single} Gaussian chain in external fields $W_A(\textbf{r})$ and $W_B(\textbf{r})$ acting on segments A and B, respectively, is
 
\begin{equation}\label{eq-qq}
  \mathcal{Q} \left[ W_A, W_B   \right] \equiv  \int  \tilde{\mathcal{D}}  \textbf{r}_{\alpha} \left( \cdotp \right) \ \exp \left[   - 
  \int_0^f ds W_A(\textbf{r}_{\alpha}(s)) - 
  \int_f^1 ds W_B(\textbf{r}_{\alpha}(s))
  \right]
\end{equation}

The path integral,  $ \int  \tilde{\mathcal{D}}  \textbf{r}_{\alpha} \left( \cdotp \right)$ , is taken over single-chain trajectories, $\textbf{r}_{\alpha} \left( s \right) $, with Wiener measure expressed as
$ \tilde{\mathcal{D}} \textbf{r}_{\alpha} =  \mathcal{D} \textbf{r}_{\alpha} P[ \textbf{r}_{\alpha}; 0, 1] $, and
\begin{equation}\label{eq2}
    P[ \textbf{r}_{\alpha}; s_1, s_2] \propto \exp \left[ - \frac{3}{2Na^2} \int_{s_1}^{s_2} ds |\frac{d}{ds}\textbf{r}_{\alpha} (s) |^2 \right]
\end{equation}
Note that $a$ is the segment size,   and $N a^2$ is the 
mean squared end-to-end distance of a Gaussian chain.
By Kac-Feynman theorem,  eq \ref{eq-qq} can be related to a Fokker-Planck partial differential equation\cite{fredrickson2006b},  known also as modified
diffusion equation (MDE)  and  shown with appropriate details  below (eqs \ref{diff1} and \ref{diff2}).

Segments $A$ and $B$ interact via the $\chi$ parameter
  which provides an effective measure of incompatibility between them\cite{deg79}. 
Evaluation of the full partition function of $n$ interacting diblock chains, shown below (eq \ref{eq1}),  is a highly
challenging task, involving many-body interactions, both intermolecular and intramolecular.
\begin{equation}{\label{eq1}}
Z = \int \prod_{\alpha = 1}^{n} \tilde{\mathcal{D}} \textbf{r}_{\alpha} \   \delta [1- \hat{\phi}_A - \hat{\phi}_B ] \ \exp\left[ -\chi \rho_0  \hat{\phi}_A \hat{\phi}_B\right] 
\end{equation}
where $\delta$-function enforces incompressibility (the melt is assumed to be incompressible), and 
\begin{equation}
\hat{\phi}_A (\textbf{r}) = \frac{N}{\rho_0} \sum_{\alpha =1}^{n} \int_0^f ds \ \delta (\textbf{r} - \textbf{r}_{\alpha} (s))
\end{equation}
\begin{equation}
\hat{\phi}_B (\textbf{r}) = \frac{N}{\rho_0} \sum_{\alpha =1}^{n} \int_f^1 ds \ \delta (\textbf{r} - \textbf{r}_{\alpha} (s))
\end{equation}
are the microscopic segments densities of $A$ and $B$, respectively; $\rho_0 = nN /V$ is the segment number density.
After replacing microscopic segment (or particle) densities with a variety of fields \cite{fredrickson2006b, mat94, ban92, vav92},
by inserting and spectrally decomposing the appropriate $\delta$-functionals, 
the partition function of an incompressible diblock melt is
\begin{equation} \label{pf1}
Z= \mathcal{N} \int \mathcal{D}\phi_A \left( \cdotp \right) \  \mathcal{D} W_A  \left( \cdotp \right) \ \mathcal{D} \phi_B \left( \cdotp \right) \  \mathcal{D} W_B  \left( \cdotp \right) \ \mathcal{D} \Psi \left( \cdotp \right)
\exp   \left[  - \frac{F\left[ \phi_A, W_A, \phi_B, W_B, \Psi   \right] }{k_BT} \right]
\end{equation}
where $\mathcal{N} $ is a normalization factor.
The functional integral is taken over the relevant fields $\phi_A\left( \textbf{r} \right), W_A\left( \textbf{r} \right), \phi_B\left( \textbf{r} \right), W_B\left( \textbf{r} \right)$, and $\Psi \left( \textbf{r} \right)$, with the free energy functional, $F\left[ \phi_A, W_A, \phi_B, W_B, \Psi   \right] $, including the single chain partition function (in external fields $W_A(\textbf{r})$ and $W_B(\textbf{r})$), as shown below 
\begin{eqnarray}
\nonumber  \frac{F \left[ \phi_A, W_A, \phi_B, W_B, \Psi   \right]}{nk_BT}  & \equiv &  - \ln \frac{\mathcal{Q}}{V} + 
   V^{-1} \int d \textbf{r}[N \chi \phi_A \left( \textbf{r} \right)  \phi_B \left( \textbf{r} \right)  \\ & &
\nonumber  - W_A \left( \textbf{r} \right) \phi_A \left( \textbf{r} \right)-   W_B \left( \textbf{r} \right) \phi_B \left( \textbf{r} \right)
 \\ & & - \Psi \left( \textbf{r} \right) ( 1 -\phi_A \left( \textbf{r} \right) - \phi_B \left( \textbf{r}) \right)] \label{free}
\end{eqnarray}
%and
%begin{equation}\label{eq-q}
 % \mathcal{Q} \equiv  \int  \tilde{\mathcal{D}}  \textbf{r}_{\alpha} \left( \cdotp \right) \ \exp\{  - 
 % \int_0^f ds W_A(\textbf{r}_{\alpha}(s)) - 
  %\int_f^1 ds W_B(\textbf{r}_{\alpha}(s))
 % \}
%\end{equation}
 Fields $\phi_A\left( \textbf{r} \right)$ and $\phi_B\left( \textbf{r} \right)$ are associated with normalized concentration profiles of $A$ and $B$, and fields   $W_A\left( \textbf{r} \right)$ and   $W_B\left( \textbf{r} \right)$ with chemical potential fields acting on $A$ and $B$, respectively;
field $\Psi \left( \textbf{r} \right)$ enforces incompressibility. Evaluating functional integrals in eq \ref{pf1} is a challenging task which, in principle,  can be performed by field theoretic simulations as proposed and implemented  by Fredrickson and coworkers\cite{fredrickson2006b, lennon2008}. A simpler, but approximate, approach is based on the mean-field idea, where the dominant, and in fact only, contribution to the functional integral in eq  \ref{pf1} comes from the fields satisfying the saddle point condition expressed as the following set of equations:
\begin{equation}\label{eq-min}
    \frac{\delta F}{\delta \phi_A} = 
    \frac{\delta F}{\delta \phi_B} = 
    \frac{\delta F}{\delta W_A} = 
\frac{\delta F}{\delta W_B} = 
     \frac{\delta F}{\delta \Psi } = 0
\end{equation}
Performing the above functional derivatives yields
\begin{eqnarray}
\label{wa} W_A(\textbf{r}) & = & N \chi \phi_B( \textbf{r}) + \Psi (\textbf{r}) \\ 
\label {wb} W_B(\textbf{r}) & = & N \chi \phi_A( \textbf{r}) + \Psi (\textbf{r}) \\
\label{psi} 1 & = & \phi_A( \textbf{r}) + \phi_B( \textbf{r})\\
\label{fia}  \phi_A (\textbf{r})& = & \frac{V}{\mathcal{Q}} \int_0^f d s \ q( \textbf{r}, s) q^\dag (\textbf{r}, s) \\
\label{fib} \phi_B (\textbf{r})& = & \frac{V}{\mathcal{Q}} \int_f^1 d s \ q( \textbf{r}, s) q^\dag (\textbf{r}, s)
\end{eqnarray}
where $\mathcal{Q}/V$ can be calculated as
\begin{equation} \label{q10}
\frac{\mathcal{Q}}{V} = \frac{1}{V} \int d \textbf{r}\  q(\textbf{r}, 1)
\end{equation}
and $q( \textbf{r}, s)$ is the forward chain propagator which is the solution of the following
modified diffusion equation
\begin{eqnarray}
 \nonumber  \frac{\partial q}{\partial s} & = & \frac{1}{6}N a^2 \nabla^2 q  - W_A(\textbf{r}) q, \ \ 0 \leq s  \leq f \\
  \frac{\partial q}{\partial s} & = & \frac{1}{6}N a^2 \nabla^2 q  - W_B(\textbf{r}) q, \ \  f \leq s \leq 1 \label{diff1}
\end{eqnarray}
with the initial condition $ q(\textbf{r}, 0)=1 $.
Similarly $q^\dag(\textbf{r}, s)$ is the  backward chain propagator which is the solution of the conjugate
modified diffusion equation:
 \begin{eqnarray}
 \nonumber  -\frac{\partial q^\dag}{\partial s} & = & \frac{1}{6}N a^2 \nabla^2 q^\dag  - W_A(\textbf{r}) q^\dag, \ \ 0 \leq s \leq f \\
  -\frac{\partial q^\dag}{\partial s} & = & \frac{1}{6}N a^2 \nabla^2 q^\dag  - W_B(\textbf{r}) q^\dag, \ \  f \leq s \leq 1 \label{diff2}
\end{eqnarray}
with the initial condition $ q^\dag(\textbf{r}, 1)=1 $. 

While the set of equations \ref{wa}, \ref{wb}, \ref{psi}, \ref{fia}, and \ref{fib} can be solved, in principle, in a self-consistent manner,
it is difficult to solve this set without some additional assumptions. 
First, we assume that the melt forms a spatially ordered nanophase.
Second, we use the UCA which is a considerable simplification, limiting  our attention to a single  D-dimensional spherical cell 
of radius $R$, and volume  $V$.  All fields, within this cell,
have radial symmetry, which  reduces this problem computationally to a single radial coordinate, $r$. 
The unconstrained spatial variables, specified in Table \ref{tab_tab4} for each $d$, become computationally irrelevant.
Thus eq \ref{q10} can be rewritten as
\begin{equation} 
\frac{\mathcal{Q}}{V} =  D \frac{\int_{0}^R r^{D-1} q(r,1) dr}{ R^D}
\end{equation}
Note that the factor, $D$,   in front of the above integral originates from the ratio of the area of a sphere with radius 1 to
the volume of a spherical cell with the same radius, both in $D$ dimensions.

While in integrals (eqs \ref{fia},  \ref{fib} and \ref{q10}) we replace $\textbf{r}$ with $r$, and $d\textbf{r}/V$ with $Dr^{D-1}dr/R^D$,  in the modified diffusion equations, \ref{diff1} and \ref{diff2}, we replace $\textbf{r}$ with $r$ and use the spherically symmetric
form of the Laplacian
 \begin{equation} \label{lap}
\nabla^2 f = \frac{\partial^2 f}{\partial r^2} +\frac{D -1}{r}\frac{\partial f }{\partial r}
\end{equation}
and similarly,  in equations for both propagators $q(r,s)$ and $q^\dagger(r,s)$, we replace $\textbf{r}$ with $r$.
Obviously the solution depends on radius, $R$, and dimensionality, $D=1, 2, 3, 4$ and 5,  corresponding
to 5 different nanophases,  shown in Table \ref{tab_tab4}.
We use the Crank-Nicholson scheme\cite{previous} to solve iteratively the modified diffusion equations (eqs \ref{diff1} and \ref{diff2}) 
 in their radial form, until the self-consistency condition is met, obtaining the saddle point 
fields,  $\overline{\phi_A} (r), \overline{\phi_B}(r), \overline{W_A}(r)$ and $\overline{W_B}(r)$  for a given $R$ and $D$. 
In the MF approximation,  the free energy functional becomes the free energy, and therefore 
 we calculate the reduced free energy (per chain in $k_BT$ units)
by substituting the saddle point fields into eq \ref{free}:
\begin{eqnarray}
\nonumber
 \frac{F(R, D)}{nk_BT}  & \equiv &  - \ln \frac{\mathcal{Q}}{V} + 
   \frac{D}{R^D} \int_0^R r^{D-1}[N \chi \overline{\phi_A}(r) \overline{\phi_B}(r)  - \\ 
& & \overline{W_A}(r) \overline{\phi_A}(r) -  \overline{W_B}(r) \overline{\phi_B}(r) ]dr \label{klklkl}
\end{eqnarray}

\newpage
\section{Results and Discussion}

Since in the  MF theory, the stability of a nanophase depends on  the product $\chi N$ and composition, $f$, 
we start, at a given point of the phase diagram, ($f$, $\chi N$),  with numerical calculation of $F(R, D)$ (eq \ref{klklkl}) for various 
$D$'s (1, 2, 3, 4, and 5) and $R$'s. 
In order to solve the MDE's (eqs \ref{diff1} and \ref{diff2}) we use up to $N_T=160$ and up to $N_R=800$ steps for the ``time'', $s$,  and space, $r$, variables, respectively.

\begin{figure}[ht]
 \includegraphics[scale=0.7]{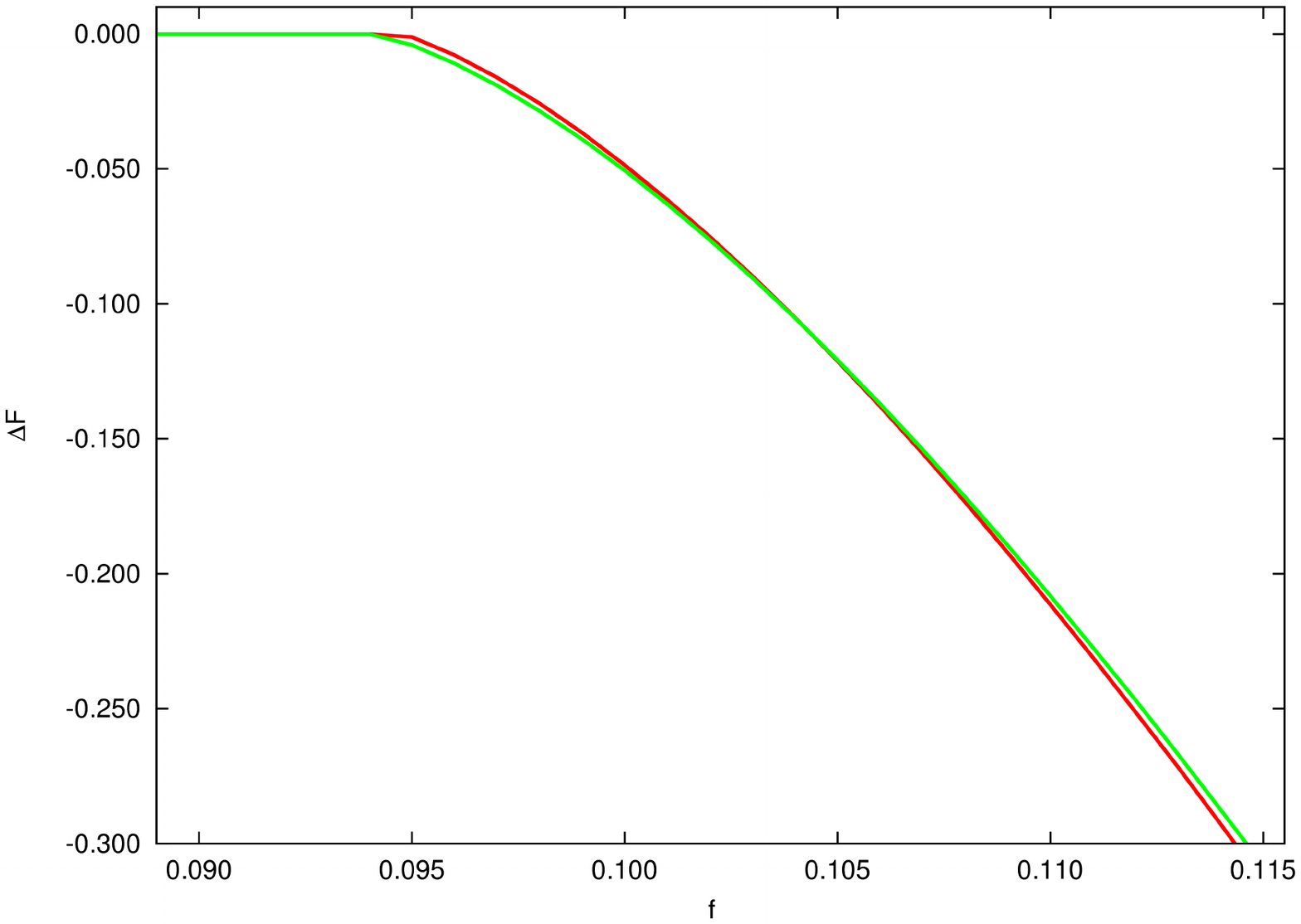} 	%pdf
 \caption{ $\Delta F$ (in $nk_BT$ units) as a function of $f$ for $\chi N =50$. Red line indicates 
the results for the $S_4$ nanostructure and green line for the $S_5$ nanostructure.
The lines intersect at $f_{S_4/S_5}$= 0.10418.
 }
\label{fig1}
 \end{figure}

Numerically, we find  $R$ and $D$ which minimize $F(R,D)$,
and this allows us to determine the dimensionality, $D$, of the most stable nanophase, and
therefore the most favorable nanophase itself,  using the correspondence from 
Table \ref{tab_tab4}.
But  the free energy of this nanophase has to be compared to 
  that of the disordered phase.
Therefore, we calculate the difference
\begin{equation}
 \frac{\Delta F}{nk_BT}\equiv  \frac{F}{nk_BT} -  \frac{F_{dis}}{nk_BT} 
\end{equation}
 where $F_{dis}$ is the free energy of the disordered phase:
\begin{equation}
 \frac{F_{dis}}{nk_BT} = N \chi f (1-f)
\end{equation}
If $ \Delta F$ is negative then the appropriate  nanophase is thermodynamically stable for the 
point considered,  ($f$,$\chi N$);  otherwise
the  system is the disordered phase. 
For example, in Figure \ref{fig1} we compare ${\Delta F}/(nk_BT)$ for $d=4$ and $d=5$ as a
function of $f$, at $\chi N = 50$. The intersection of those free energy curves occurs at 
$f_{S_4/S_5} = 0.10418$, as also indicated in Table \ref{tab:res}.

%----------------------------------------------------------------------

\begin{figure}[ht]
 \includegraphics[scale=0.7]{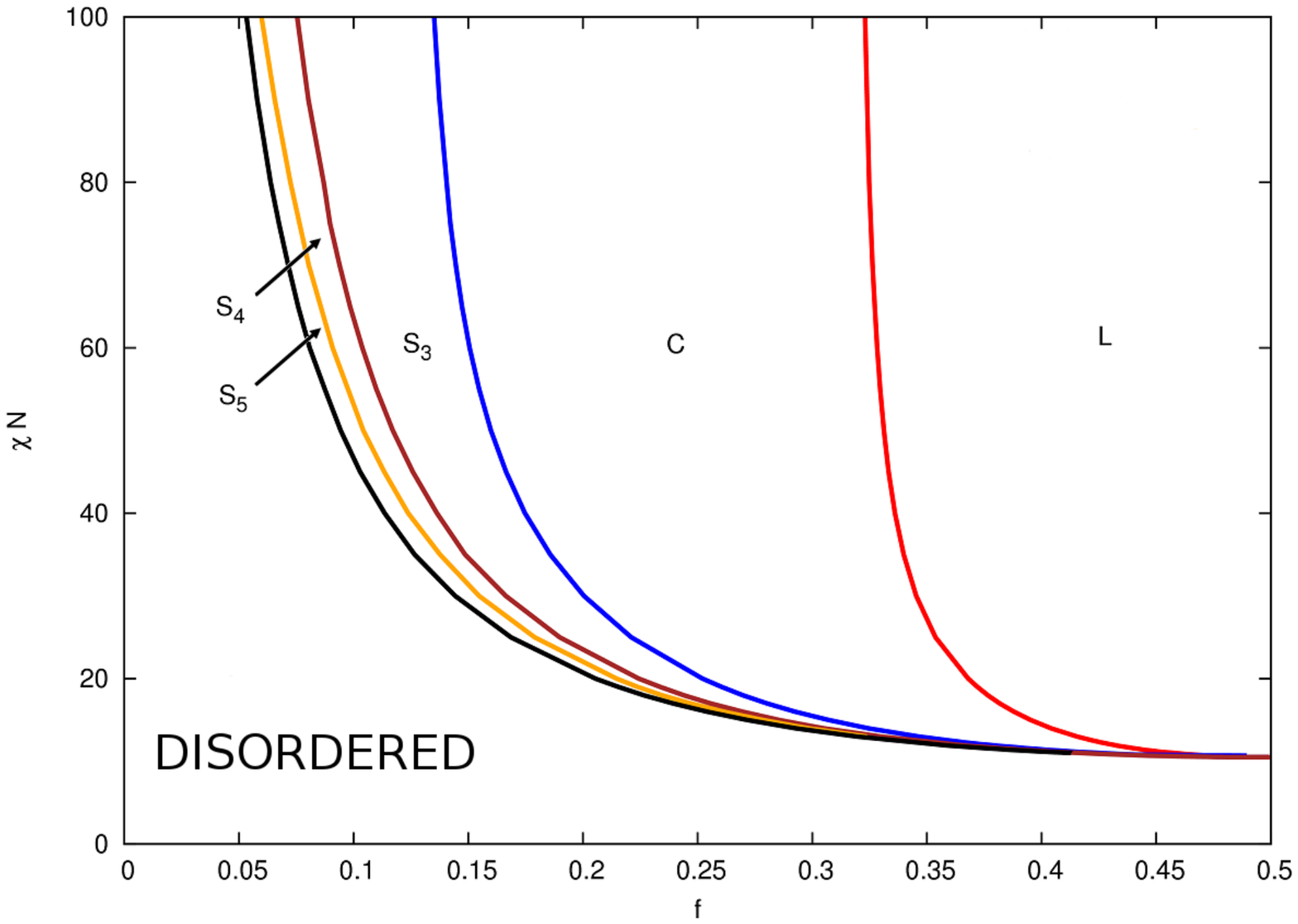} 	%pdf
 \caption{DBC phase diagram in 5$d$: $L$, $C$, $S_3$,  $S_4$, and $S_5$ indicate corresponding nanophases;  the disordered
phase is also shown.
 }
\label{fig2}
 \end{figure}

This procedure allows us to map the DBC melt phase diagram for  $d=5$
 in the $(f, \chi N)$-plane, as shown in Fig \ref{fig2}.
Since there is a mirror symmetry with respect to $f=0.5$ ($f \rightarrow 1-f$, $A$ can be exchanged with $B$), we show  the resultant nanophases only from $f=0$ to 0.5, and 
the following phase sequence  is observed: $L$, $C$, $S_3$,  $S_4$, $S_5$ and the disordered phase; the corresponding data for those lines is presented in Table \ref{tab:res}. 
A new nanophase, $S_5$,  is observed  in a relatively narrow strip
between the $S_4$ phase and disordered phase. This is the main result of this paper. 
We extrapolate the calculated OOT lines, $f_{L/C}$, $f_{C/S_3}$, $f_{S_3/S_4}$, $f_{S_4/S_5}$to 
the strong segregation limit, that is we estimate them as $\chi N \rightarrow \infty$ (or 1/$(\chi N) \rightarrow 0$ ),
fitting  to the following function:
\begin{equation}
f(\chi N)= f ^0+\frac{g^0}{\chi N}
\label{fit111}
\end{equation}
as used in referenced \cite{whit96} and \cite{previous}, where $f^0$ is the extrapolation to the strong segregation limit, and $g^0$ is a fitting parameter.   
The resultant limits, $f_{L/C}^0$, $f_{C/S_3}^0$, $f_{S_3/S_4}^0$, as determined in ref \onlinecite{previous} and $f_{S_4/S_5}^0$ (determined in this paper)
are  compared to the SST $f$'s, as shown in  
Table \ref{tab:res1}. 
The discrepancy between the SST and the present SCFT with the UCA, for  $f_{L/C}$ and $f_{C/S_3}$  is
within 2\% error, as also reported in \cite{whit96}, and the discrepancy for $f_{S_3/S_4}$ is about 10\%.
However, the difference between the extrapolated $f^0_{S_4/S_5}=0.01$ and the calculated $f_{S_4/S_5}=0.00018$ (from SST) is much larger. Since the SCFT is more advanced and accurate
theory than the SST (the SCFT is a full mean field theory, and the SST is an approximation of the mean field theory which is meaningful only at strong segregations), we demonstrate that the area of stability for the $S_5$ phase is mostly likely
to be siginicantly larger than that predicted from the SST.

\begin{figure}[ht]
 \includegraphics[scale=0.75]{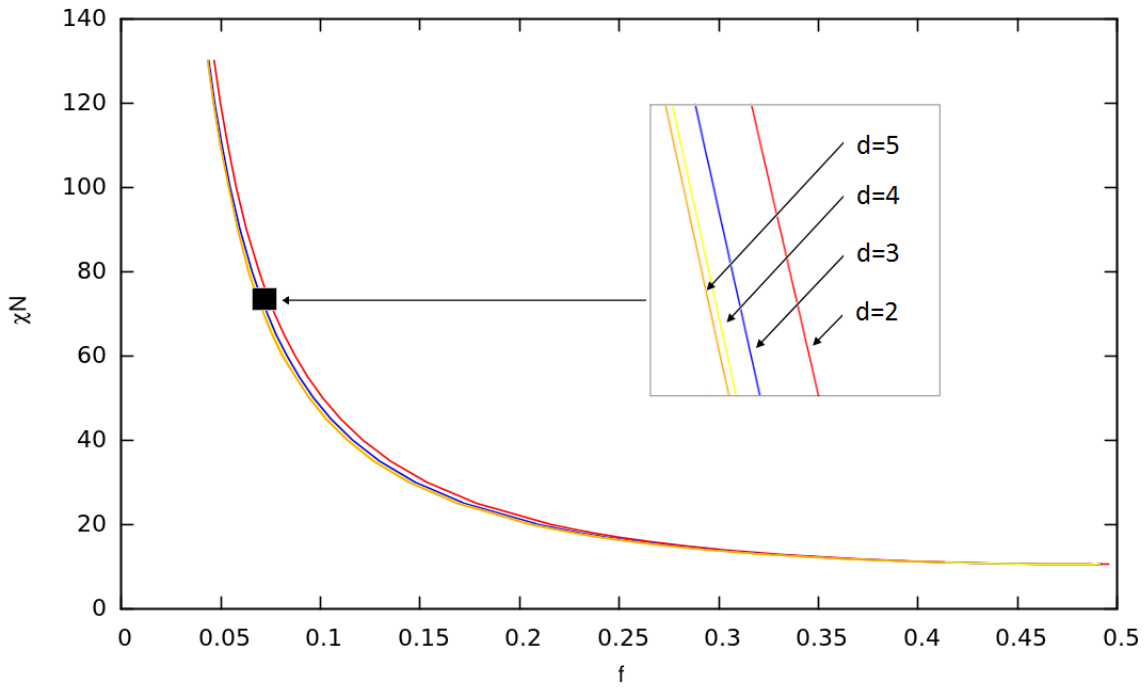} 	%pdf
 \caption{The ODT lines for $d$=2, 3, 4 and 5. The inset is from $\chi N = 70$ to 75.
 }
\label{fig3}
 \end{figure}

While spinodals for the ODT calculated with random-phase approximation (RPA) \cite{lei80:9} are the same for $d=2$, 3,  4,  and 5 the binodals (the ODT lines), calculated in this work,  depend on $d$ as shown in Fig \ref{fig3}. The binodals depend  weakly on $d$, and they are particularly close to each other in the 
vicinity of  $f_A = 1/2$ (symmetric diblock), and
therefore we show them in a narrow window (from 70 to 75 in $\chi N$, that is away from $f=1/2$)
in the inset of Fig \ref{fig3}. We observe the sequence of binodals, as shown in inset of Fig \ref{fig3}. 
For $f=1/2$ the RPA spinodal is at $(\chi N)_c \approx 10.4949$, and the calculated binodals (for $d$=2, 3, 4 and 5)
also  converge to this point within  the numerical accuracy. Similarly, the OOT lines seem to
converge to  $(\chi N)_c$ for  $f=0.5$.

\begin{table}[t]
\centering
\begin{tabular}{|c|c|c|c|c|c|}
\hline
$\chi N$ & $f_{L/C}$ & $f_{C/S_3}$ & $f_{S_3/S_4}$ & $f_{S_4/S_5}$ & $f_{ODT}$ \\ \hline
20 & 0.36797 & 0.25210  & 0.22436 & 0.21461 & 0.20541   \\ \hline
30 & 0.34531 & 0.20048  & 0.16643 & 0.15475 & 0.14434   \\ \hline
40 & 0.33601 & 0.17474  & 0.13631 & 0.12379 & 0.11347   \\ \hline
50 & 0.33120 & 0.15988  & 0.11714 & 0.10418 & 0.09439  \\ \hline
60 & 0.32828 & 0.15059  & 0.10379 & 0.09086 & 0.08086 \\ \hline
70 & 0.32629 & 0.14452  & 0.09387 & 0.08043 & 0.07143 \\ \hline
80 & 0.32482 & 0.14057  & 0.08698 & 0.07143 & 0.06383 \\ \hline
90 & 0.32392 & 0.13737  & 0.08029 & 0.06556 & 0.05789\\ \hline
100 & 0.32303& 0.13510  & 0.07535 & 0.05983 & 0.05332\\ \hline
\end{tabular}
\caption{The ODT and OOT lines for selected $\chi N$'s}
\label{tab:res}
\end{table}

\begin{table}
\caption{The OOT lines from the full SCFT\cite{whit96} and UCA extrapolated to infinite $\chi N$'s compared 
to the SST results}
\label{tab:res1}
\centering
\scriptsize
\begin{tabular}{|c||c|c|c|c|c|}
\hline
 Method &\ \ \  $f^0_{L/C}$ \ \ \  & \ \ \ $f^0_{C/S_3}$ \ \ \  &\ \ \  $f^0_{S_3/S_4}$ \cite{previous} \ \ \  &\ \ \  $f^0_{S_4/S_5}$ \ \ \  \\ \hline \hline

full SCFT \cite{whit96} & 0.3100 & 0.1050 & -  & -  \\ \hline

\textbf{UCA} &  \textbf{0.3150}  & \textbf{0.1149}  &  \textbf{0.0306} & \textbf{0.010 }\\ \hline

\textbf{SST}    &  \textbf{0.2999} &\textbf{0.1172} &  \textbf{0.0336} & \textbf{0.00018} \\ \hline

\end{tabular}
\end{table}

\section{Conclusions}

Using a self-consistent field theory in spherical unit cells of various dimensionalities, $D$=1, 2, 3, 4 and 5, we calculate  phase diagram of a diblock, $A$-b-$B$,
copolymer melt in 5-dimensional space, $d$=5.
The phase diagram is parameterized by the chain composition, $f$, and  incompatibility between $A$ and $B$,  quantified by the product $\chi N $.
We predict 5 stable nanophases:
layers, cylinders, 3$D$ spherical cells, 4$D$  spherical cells  and 5$D$  spherical cells, and  calculate both order-disorder and
order-order transition lines. In the strong segregation limit, that is for large $\chi N$, the OOT  compositions,  $f_{L/C}$,  $f_{C/S_3}$, $f_{S_3/S_4}$ and $f_{S_4/S_5}$ are determined by the strong segregation theory. 
While $f_{L/C}$,  $f_{C/S_3}$, and $f_{S_3/S_4}$   are close   to the
corresponding extrapolations from the self-consistent field theory, as shown known
in the previous study \cite{previous}, the $f_{S_4/S_5}$ extrapolation does not agree with the SST predictions.
We find that 
 the $S_5$ nanophase is stable in a narrow strip between
ordered $S_4$ nanophase and the disordered phase.
The calculated binodals (ODT lines) depend weakly on $d$, as expected.

\begin{acknowledgments}
We acknowledge a computational grant from the Poznan Supercomputing and Networking Center (PCSS).
\end{acknowledgments}

\newpage
\bibliography{scft_first}

\newpage

% figure microarchitectures

\end{document}